\documentclass[mnsc,nonblindrev]{informs3}

\OneAndAHalfSpacedXI

\usepackage{natbib}
 \bibpunct[, ]{(}{)}{,}{a}{}{,}%
 \def\bibfont{\small}%

\usepackage[normalem]{ulem}
\usepackage[utf8]{inputenc} 
\usepackage[T1]{fontenc}    
\usepackage{hyperref}
\hypersetup{colorlinks,allcolors=black}       
\usepackage{url}            
\usepackage{booktabs}       
\usepackage{amsfonts}       
\usepackage{nicefrac}       
\usepackage{microtype}      
\usepackage{lipsum}
\usepackage{graphicx}
\usepackage[subtle]{savetrees}

\usepackage{placeins}
\usepackage{color}
\usepackage{authblk}
\usepackage{natbib}
\usepackage{amsmath}
\usepackage[title]{appendix}
\usepackage{setspace}
\usepackage[margin=1in]{geometry}

\definecolor{mgcol}{rgb}{0.8, 0.2, 0.8}

\definecolor{mocol}{RGB}{150, 80,150}

\definecolor{brown}{RGB}{165,42,42}

\begin{document}

\RUNAUTHOR{Carranza, Goic, Lara, Olivares,  Weintraub, et.al.}

\RUNTITLE{The Social Divide of Social Distancing}

\TITLE{The Social Divide of Social Distancing: Shelter-in-Place Behavior in Santiago during the Covid-19 Pandemic\footnote{Carranza, Goic, Lara, Olivares, and Weintraub are the main contributors of this work and listed in alphabetical order. Covarrubia, Escobedo, and Jara built the raw mobility data. Basso is the director of ISCI, which hosted the project, and also provided valuable input to the results and writing.}}

\ARTICLEAUTHORS{
\AUTHOR{Aldo Carranza}
\AFF{Stanford University, \EMAIL{aldogael@stanford.edu}} 
\AUTHOR{Marcel Goic}
\AFF{Ingeniería Industrial, Universidad de Chile and Instituto Sistemas Complejos de Ingenieria (ISCI), \EMAIL{mgoic@uchile.cl}}
\AUTHOR{Eduardo Lara}
\AFF{Ingeniería Industrial, Universidad de Chile and Instituto Sistemas Complejos de Ingenieria (ISCI), \EMAIL{eduardo.lara@ug.uchile.cl}}
\AUTHOR{Marcelo Olivares}
\AFF{Ingeniería Industrial, Universidad de Chile and Instituto Sistemas Complejos de Ingenieria (ISCI), \EMAIL{molivares@uchile.cl}}
\AUTHOR{Gabriel Y. Weintraub}
\AFF{Stanford Graduate School of Business, Stanford University,
\EMAIL{gweintra@stanford.edu}}
\AUTHOR{Julio Covarrubia}
\AFF{Digital Entel Ocean, Empresa Nacional de Telecomunicaciones (ENTEL)}
\AUTHOR{Cristian Escobedo}
\AFF{Digital Entel Ocean, Empresa Nacional de Telecomunicaciones (ENTEL) and Facultad de Arquitectura y Urbanismo, Universidad de Chile}
\AUTHOR{Natalia Jara}
\AFF{Digital Entel Ocean, Empresa Nacional de Telecomunicaciones (ENTEL)}
\AUTHOR{Leonardo J. Basso}
\AFF{Ingeniería Civil, Universidad de Chile and Instituto Sistemas Complejos de Ingenieria (ISCI),\EMAIL{ljbasso@gmail.com}}
} 



\maketitle

Voluntary shelter-in-place directives and lockdowns are the main non-pharmaceutical interventions that governments around the globe have used to contain the Covid-19 pandemic. In this paper we study the impact of such interventions in the capital of a developing country, Santiago, Chile, that exhibits large socioeconomic inequality. A distinctive feature of our study is that we use granular geolocated mobile phone data to construct mobility measures that capture (1) shelter-in-place behavior, and (2) trips within the city to destinations with potentially different risk profiles.
Using panel data linear regression models we first show that the impact of social distancing measures and lockdowns on mobility is highly heterogeneous and dependent on socioeconomic levels. More specifically, our estimates indicate that while zones of high socioeconomic levels can exhibit reductions in mobility of around 50\% to 90\% depending on the specific mobility metric used, these reductions are only 20\% to 50\% for lower-income communities. The large reductions in higher-income communities are significantly driven by voluntary shelter-in-place behavior. Second, also using panel data methods we show that our mobility measures are important predictors of infections: roughly, a 10\% increase in mobility correlates with  a 5\% increase in the rate of infection. Our results suggest that mobility is an important factor explaining differences in infections rates between high and low incomes areas within the city. 
Further, they confirm the challenges of reducing mobility in lower-income communities, where people generate their income from their daily work. To be effective, shelter-in-place restrictions in municipalities of low socioeconomic levels may need to be complemented by other supporting measures that enable their inhabitants to increase compliance.  






\vspace{5mm}

\section{Introduction}
In the face of the global Covid-19 crisis, governments around the world have rushed to design and implement mitigation strategies to contain the pandemic. Most countries have continuously asked the population to practice physical distance -- in particular, to shelter in place -- and have adopted more radical measures of mandatory lockdowns
to  flatten the curve of infections and prevent a collapse of the healthcare system \citep{hsiang2020effect}. However, the effectiveness of the adopted measures in terms of mitigating the spread of the virus may vary significantly across the population groups where they are implemented \citep{allcott2020polarization,lamb2021differential}. Our work seeks to identify factors that help to explain the heterogeneous effects of social distancing  in the capital of  a developing country:  the city of Santiago, Chile, with a population of around 7 million. With this evidence we shed light on general policy prescriptions to contain the pandemic in large urban cities that exhibit significant socioeconomic  disparities \citep{florida2017new}.

\begin{figure}[htb]
    \centering
    \includegraphics[width=0.7\textwidth]{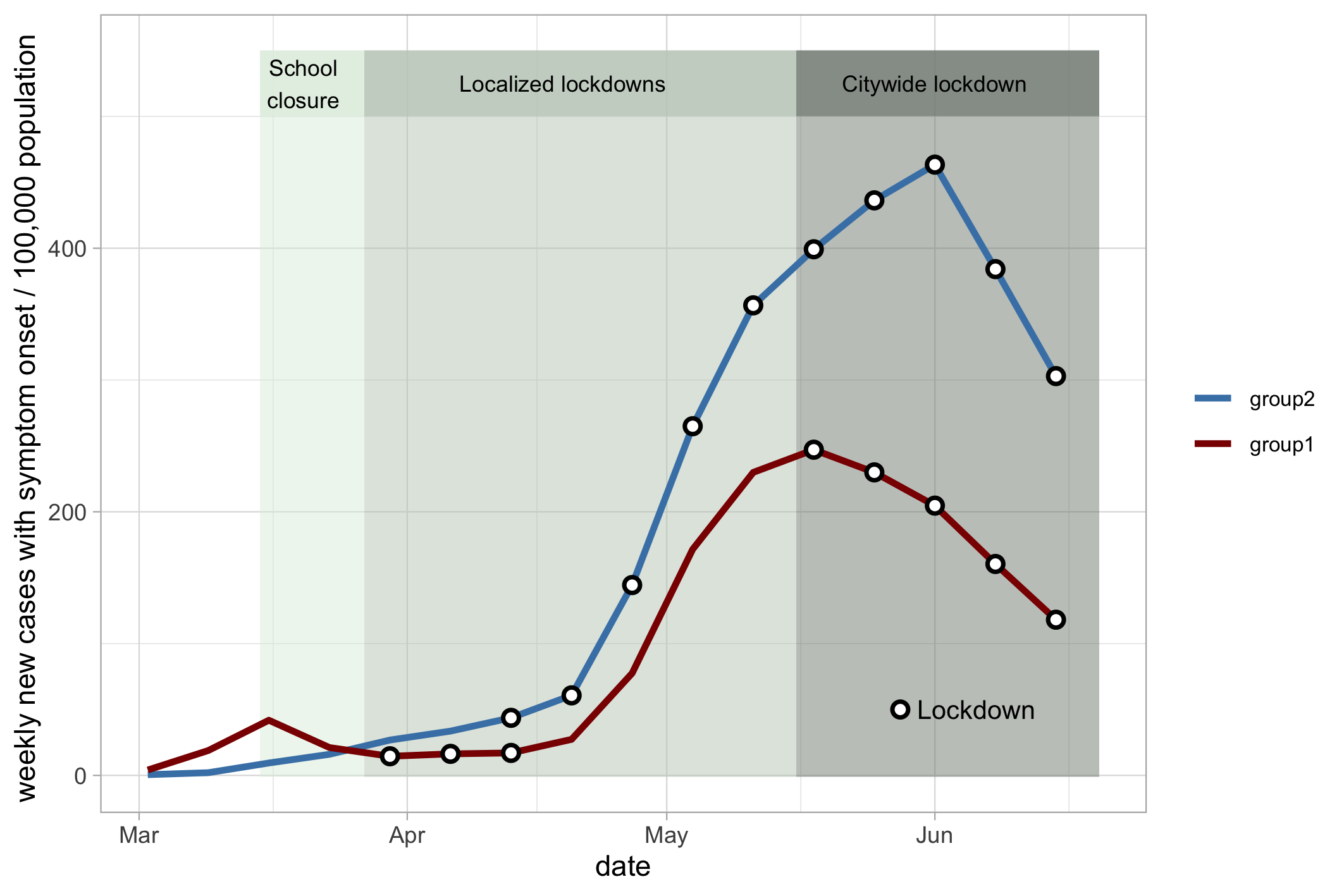}
    \caption{New cases by week of symptom onset, grouped by areas where the first and second waves of localized lockdowns in Santiago were implemented.}
    \label{fig:cases_group}
\end{figure}

Chile's first case, a traveler from Singapore, was detected on March 3, 2020.  The date coincides with the start of the working and academic year (January and February correspond to school summer vacations). The virus spread rapidly, with an average doubling time of 2--3 days in the first few weeks. On March 15, the government mandated the closure of all schools in the country, and called on the public to implement social distancing and remote work as much as possible. On March 30, the government issued the first lockdown order in Santiago, but included only the municipalities in the center and eastern part of the city that have an approximate population of 1.2 million. Schools continued to be closed and social distancing was recommended for the entire city. 
Figure \ref{fig:cases_group} shows the number of new Covid-19 cases by week of symptom onset for two groups representing different geographic areas of Santiago; dots in the figure indicate weeks when the localized lockdown was in place. Localized lockdowns in different parts of the city were enforced through mid-May, after which a citywide lockdown was implemented.\footnote{Some municipalities  were only partially locked down but cases are reported for the whole municipality. Group 1 includes the municipalities located in the eastern part of the city, where the initial outbreak originated and the first lockdown was implemented. The voluntary shelter-in-place directive and the mandated localized lockdown appeared to stabilize the outbreak in this group. Consequently, the lockdown of these municipalities was lifted around mid-April, followed by a second lockdown of other areas of the city with growing cases, labeled as group 2. Unfortunately, the containment of the outbreak was much less effective in group 2, with a steady increase in the number of new cases, higher even than in group 1 which was not under lockdown. This led to an explosion of new cases in mid-May and the government mandated a citywide lockdown for Santiago on May 15. Cases continued to grow through mid-June, with a near collapse of critical-care beds in Santiago -- in spite of more than doubling the ICU capacity \citep{goic2021covid}.}

Why was it harder to contain the pandemic in group 2? 
Our work aims to understand how voluntary shelter-in-place behavior and compliance to lockdowns may vary substantially depending on the characteristics of the population in a city. 
In fact, there are important differences in shelter-in-place behavior between groups, with group 1 reducing its mobility significantly more than group 2. 
Furthermore, it is in fact revealing that 61\% of the total population in group 1 belongs to the highest-income segment. This is in part because the pandemic was mostly initiated by foreign travelers and the initial outbreak was concentrated in the higher-income areas. By contrast, only 2.5\% of group 2 belongs to the  highest-income segment. More generally, in developing countries like Chile there is typically high heterogeneity in socioeconomic and educational levels within cities.\footnote{The Metropolitan Region that includes the city of Santiago has a Gini index of 0.5, the largest in the country, and exhibits not only important income inequalities, but also high levels of educational and urban segregation \citep{romero2012assessing, elacqua2012impact}.} 


This analysis provides evidence that the interaction between mobility and socioeconomic status is important for understanding the spread of the pandemic. In this study we show how people's socioeconomic level has a significant impact on people's response to voluntary shelter-in-place directives and lockdowns, the most widely used non-pharmaceutical interventions in the Covid-19 pandemic.
A distinctive feature of our study is that we consider granular location data to characterize how mitigation measures reduce the population's mobility and estimate how these reductions contain the rate of infection. More specifically, to study mobility we collected detailed data from the usage of the wireless telecommunications infrastructure to build localized measures of mobility, capturing shelter-in-place adherence and population flows between local areas. 

While other studies also use mobility data to study the effect of lockdowns, our approach is novel in terms of specifically measuring trips and shelter-in-place behavior within a specific city. This is in contrast to capturing heterogeneous behavior also with respect to income but across an entire country such as the U.S. \citep{weill2020social}. Also at the level of the U.S., \cite{chiou2020social} explores the role of not just income but also internet access in adherence to shelter-in-place directives. In addition, \cite{valentino2020location} makes a model-free comparison of mobility between the wealthiest and the poorest areas in several metropolitan areas in the U.S. Finally, \cite{bonaccorsi2020economic} uses Facebook data to analyze the impact of lockdowns on mobility in Italy.
In contrast to these studies, our measures of lockdown compliance are localized, measuring mobility in geographic units of 2,000--3,000 people, which provides a sample size of more than 1,600 zones in the city of Santiago during a four-month period. This level of granularity within a city allows us to determine not only whether people leave their residences, but also whether they are traveling to riskier areas of the city and are more likely to interact with the infected population. Building these risk-adjusted mobility measures is an important contribution of our work. 
 
In more detail, our first main contribution is a detailed analysis to identify the causal effects of lockdowns and voluntary shelter-in-place directives on mobility and how these effects are moderated by demographics. Our identification strategy takes advantage of the localized implementation of lockdowns and the heterogeneity of the demographic characteristics across the city. This analysis with panel data reveals that socioeconomic characteristics are key in determining voluntary shelter-in-place behavior and lockdown compliance.
Our estimates indicate that while zones of high socioeconomic levels can exhibit reductions in mobility of around 50\% to 90\% depending on the specific mobility metric used, these reductions are only 20\% to 50\% for lower-income communities. The large reductions in higher-income communities are significantly driven by voluntary shelter-in-place behavior.
These differences that we found for Santiago, a metropolitan city in a developing country, are comparable in magnitude to those reported in previous work across an entire developed country \citep{weill2020social}. 

Using panel data methods, we then show that these constructs of mobility are an important predictor of outbreaks at the municipality level. We show that a 67\% reduction in mobility -- which is about the difference in mobility observed between high- and low-income municipalities  -- correlates with a 36\% reduction in the rate of infection (an elasticity of 0.5 approximately).  This result contributes to the study of the impact of non-pharmaceutical interventions on epidemiological outcomes \citep{flaxman2020estimating,villas2020we,fang2020human,allcott2020economic}. 

Our work is closely related to \cite{cuadrado2020impact}, which studies the impact of lockdowns on mobility and infections in specific regions of Chile. Unlike their work, our study focuses on a single city (Santiago) that was subject to localized lockdowns, using a more granular analysis to identify the moderating effect of demographics on the response to lockdowns and voluntary shelter-in-place directives. Our work is also related to \cite{bennett2021all},  which uses a synthetic control method to analyze the impact of lockdowns in the city of Santiago, showing that socioeconomic conditions influence their effect on the rate of infection significantly. Whereas in that study the focus is on the heterogeneous impact of lockdowns on infections, our work is focused in identifying mobility as the main underlying mechanism;  in this sense, these two studies are complementary. 

Our results show that  failing to acknowledge the heterogeneous effect of voluntary shelter-in-place directives and lockdowns can have dramatic consequences in the containment of the pandemic. It also confirms the challenges of reducing mobility in lower-income communities, where people generate their income from their daily work and cannot work remotely. Finally, our results suggest that, to be effective, shelter-in-place restrictions in municipalities of low socioeconomic levels might have to be complemented by other measures that induce their inhabitants to increase compliance. Given that these communities do not have easy access to basic services, it seems fundamental to provide aid to cover basic needs.
 
Some of the results of this study were made publicly available during the first weeks of the citywide lockdowns, receiving substantial attention from the public press and government authorities, underscoring the importance of reducing mobility to control the pandemic and of providing aid to low-income families to help them adhere to the lockdown and shelter-in-place requirements.\footnote{Please refer to https://isci.cl/tag/geointeligencia/.}

\section{Data and measures of mobility}

This study uses four different data sources for Santiago, Chile: (i) Covid-19 infections, (ii) dates of government-imposed lockdowns, (iii) demographic information, and (iv) mobility data from mobile phones. For our analysis, we selected data from March 2 to July 3, 2020, in order to include data from points in time before the first case of Covid-19 was detected in Santiago, during partial lockdowns of the city, and during total lockdowns  of the city. In addition, we considered only the urban municipalities of Santiago, which number 36 and together have a population of 6.1 million people. 

\subsection{Infections} 

Information on Covid-19 infections were  collected from the official public data made available by the Ministry of Health of Chile based on confirmed PCR tests (see \textit{Data Availability} in Appendix). We used data on the number of cases of infected individuals reported by date of initial symptoms (the only part of case counts that is self-reported is the date of symptom onset). The data is available weekly at the municipality level. Let $y_{\ell t}$ be the total number of infections in municipality $\ell$ in week $t$. In Appendix E we discuss a robustness analysis to check for underreporting of cases due to testing capacity and other factors.

\subsection{Mitigation measures} 

Information on the mitigation measures that were implemented in the city of Santiago was collected through press releases documenting the date of school closures and social distancing recommendations. Localized and citywide lockdowns were obtained from public data sources published by the Chilean Ministry of Science (see \textit{Data Availability}). These data are at the municipality level. Throughout the period, voluntary sheltering in place was strongly urged by the government and media.

\subsection{Demographics} 

Demographic information was collected from two different sources. The first source is the 2017 Chilean Census. This information is  available at the \emph{census unit} level and includes total population, elderly population (over 65 years), immigrant population, average household size, household density (average number of households per $km^{2}$), among other variables. For urban areas, the census unit corresponds to a city \emph{block}, which is the minimum aggregation unit we observe for Santiago. 

The second source of demographic information is a public data set from 2012, which provides the predominant socioeconomic group (SEG) at the block level. SEGs are based on income, education, and employment of the head of household \citep{meier2004social}. Following a common practice in other studies  \citep{romero2012assessing, hernandez2018twelve}, we grouped SEGs into three categories: high, medium, and low socioeconomic groups (HighSEG, MedSEG, and LowSEG, respectively). These demographics were aggregated to the census zone level. Each census zone is composed of several census blocks and has an average population of 3,661. The city of Santiago has 1,864 census zones, with 1,639 of them contained within the urban municipalities. Using principal component and cluster analysis we show that the socioeconomic classification defined above (HighSEG, MedSEG, and LowSEG) is highly correlated with all other demographic indicators (see Appendix A for details). Hence, for the econometric analysis we use these socioeconomic groups as our main variables of demographics. 

\subsection{Mobility data} \label{sec:mobility}

Mobility data was collected via anonymized mobile phone usage, georeferenced by triangulation of antenna connections \citep{drane1998positioning}. Although other studies have also used measures of mobility based on mobile phone usage (e.g, \cite{GoogleMobilityReport, SafeGraph}), for this study we developed new  measures capturing both origins and destinations of mobility patterns in the city. These are specifically designed to measure shelter-in-place behavior, compliance with lockdowns, as well as population movements throughout the city.

Each mobile phone is associated with a \textit{home zone}, defined as the census zone in which the device is most frequently observed between 9--11pm during the month of observation. To prevent spurious associations with home zones, in the process we considered only users with a minimum of five observations in that location during the month. On average, in each census zone we observe 53 devices per every 100 inhabitants. Considering that there is an estimated 1.35 devices per capita, our sample seems to be quite large, capturing on average more than a third of the devices, while the rest would be from different providers. Furthermore, since the value proposition of different providers is quite similar we believe our sample of mobile phones is representative of the population. There is little product differentiation in the market and based on our exploration of the data, the participation of our mobile provider was slightly larger for higher socioeconomic groups. Hence, if any bias exists, it would lean towards underestimating the difference in mobility between high- and
low-income zones (since we are likely to over sample the relatively higher-income population from the low-income zones).


To measure mobility, the location of each (anonymous) device was tracked on each day during two non-overlapping time intervals: AM (10am--1pm) and PM (2--5pm). These time intervals were chosen to capture work-commuting patterns and, consequently, the sample excludes weekends and holidays.  For each time interval on each day, every observed device is assigned to  a \textit{destination zone} defined as the census zone in which the device is most frequently observed. This information gives us an anonymized description of the mobility of the population. 

\subsection{Mobility measures} \label{sec:mob_meas}

The information about movements of devices from their home zone to other zones in the city provides an origin-destination matrix that is constructed as follows.  Define $\hat f^b_{ijd}$ as the number of devices from home zone $i$ that traveled to destination zone $j$ during the time interval $b\in\{AM,PM\}$ on day $d$.\footnote{The sum $\sum_j \hat f^b_{ijd}$ represents the number of devices from home zone $i$ that were observed in each day and interval.} Then, the weekly average number of devices from home zone $i$ observed in destination zone $j$ in week $t$ in each time interval $b$, $f_{ijt}^b$, is calculated as the mean of the daily values excluding weekends and holidays in this aggregate. Furthermore, to have a conservative measure that avoids double-counting of individuals that may appear in both time intervals, we extract a single weekly matrix by taking the entry-wise maximum between both matrices, $f_{ijt}=\max\{f_{ijt}^{AM}, f_{ijt}^{PM}\}$. For robustness, we also study a specification considering the sum of $f_{ijt}^{AM}$ and $f_{ijt}^{PM}$, that is, $f_{ijt}=f_{ijt}^{AM} + f_{ijt}^{PM}$ (see Appendix E). The results obtained from both specifications are similar. 

Under the assumption that the sample of devices is a representative sample of the population, the measure $f_{ijt}$ is a proxy for the weekly average number of people who leave their home zone $i$ for destination zone $j$ in week $t$. Using this information, we calculate measures of outward mobility in each zone:
\begin{align}
  \bar f_{ijt}=\frac{f_{ijt}}{\sum_{j}f_{ijt}}
 \quad & \mbox{and} \quad
MobOut_{it} =\sum_{j:j\neq i} \bar f_{ijt} 
\end{align}
Thus, $\bar f_{ijt}$ is a proxy for the fraction of the weekly average number of people with residences in zone $i$ who travel to zone $j$ in week $t$, while $MobOut_{it}$ is a proxy for the fraction of the weekly average number of people with residences in zone $i$ who leave that zone in week $t$. 

While the analysis of the impact of lockdowns on mobility is done at a granular census zone level, to conduct the analysis of the effect of mobility on infections, we aggregate this zone-level mobility measure to the municipality level via a population-weighted aggregation. This is because municipalities are the most granular units for which infection data is available. With some abuse of notation we define the counterpart of $MobOut_{it}$, but at the municipality level, as
\begin{align}
 MobOut_{kt}
=\frac{\sum_{i:c(i)=k}\sum_{j:j\neq i}Pop_i\cdot \bar f_{ijt}}{\sum_{i:c(i)=k}Pop_i}\ , \label{eq:MobOut}
\end{align}
where $c(i)$ denotes the municipality containing census zone $i$ and $Pop_i$ denotes the population size of census zone $i$. This measure corresponds to the proportion of the population of municipality $k$ that leaves its home zone, and so provides an aggregated measure of outward mobility.\footnote{The data reveals that the number of mobile phones observed in each home zone typically decreases with time in our sample. One potential explanation is that devices used at home use WiFi connections more frequently, reducing the amount of cellular data and thereby making these devices less likely to appear in the sample. Ignoring this effect may lead to underestimating shelter-in-place compliance. Hence, we developed a correction that we also use in some of our analysis, explained in Appendix E.}


Finally, we define a more granular mobility measure that accounts for the infection load of different areas of the city, to capture risk of infection to which people are exposed when traveling across the city. To this end, following the same logic as \eqref{eq:MobOut},
we define a population-weighted outward mobility proportion from municipality $k$ to municipality $\ell$ in week $t$ as
\begin{align*}
    \tilde f_{k\ell t}=\frac{\sum_{i:c(i)=k}\sum_{j:c(j)=\ell, j\neq i}Pop_i\cdot \bar f_{ijt}}{\sum_{i:c(i)=k}Pop_i}.
\end{align*}
Recall that $y_{lt}$ is the total number of infections in municipality $\ell$ in week $t$ and $\sum_{\ell}  y_{\ell t}$ is the total number of infections in the city. Based on this data, we define an infection-risk measure for municipality $m$ in week $t$, where we multiply by 100 to re-scale the risk metric appropriately: 
\begin{align*}
     Risk_{mt}=100\cdot\frac{\sum_{\ell}\tilde f_{\ell mt}\cdot y_{\ell t}}{\sum_{\ell}y_{\ell t}} ,
\end{align*}
which represents the proportion of infected individuals in the city that enter municipality $m$ in week $t$. This is used to construct a risk-weighted mobility measure for municipality $k$ in week $t$:
\begin{align*}
    MobRisk_{kt}=\sum_{m}\tilde f_{kmt}\times Risk_{mt}  
\end{align*}
which is a proxy for the risk that an individual from municipality $k$ will be exposed to an infected individual when traveling across the city. \textit{MobRisk} weights the outflows of the focal municipality ($k$) with the risk of the destination municipality ($m$), and sums over all municipalities, thereby accounting for the externality generated by the mobility of potentially infected people across the city.

Similarly, we define a measure of mobility risk at the census zone level. Let
$\hat f_{i\ell t}=\sum_{j:c(j)=\ell, j\neq i}\bar f_{ijt} \ ,$ which is a proxy for the fraction of the weekly average number of people with homes in zone $i$ who travel to municipality  $\ell$ in week $t$. This is used to construct a risk-weighted mobility measure for each census zone (with some abuse of notation):
$
    MobRisk_{it}=\sum_{m}\hat f_{imt}\times Risk_{mt}.  
$

In Appendix D we validate our measures and in Appendix E we also consider alternative measures of mobility using data from the integrated public transport system of Santiago.

\subsection{Descriptive analysis of mobility}

The left panel of Figure~\ref{fig:Temporal_evolution_of_mobility} shows the temporal evolution of $MobOut_{kt}$ for different municipalities of Santiago that 
represent different socioeconomic strata. The measure is normalized with respect to a baseline period that  corresponds to March 2--13, which were the first weeks of the epidemic in Chile, when it was still confined to a few cases and mobility had minimal variation. The points mark weeks where the municipality had more than half of its population under lockdown.\footnote{By construction, the first weeks taken as a baseline have essentially zero variation.}
From the week of March 16 we begin to observe a marked decrease in mobility; this date coincides with the closing of schools in Chile. Despite the fact that the schools closed concurrently throughout the city of Santiago, the effect of voluntary shelter-in-place behavior is very heterogeneous. For higher-income municipalities such as Las Condes, mobility decreased by 30\% and the following week by 40\%. By contrast, the effects on mobility were between 10\% and 20\% for lower-income districts such as Puente Alto, and even less for El Bosque. The effects for the municipalities of Santiago and Ñuñoa are somewhere in between.\footnote{We note that the fraction of HighSEG in Las Condes is 67\%, while for Ñuñoa is only 28\%.}  
\textbf{\begin{figure}[htb]
    \centering
    \includegraphics[width=\textwidth]{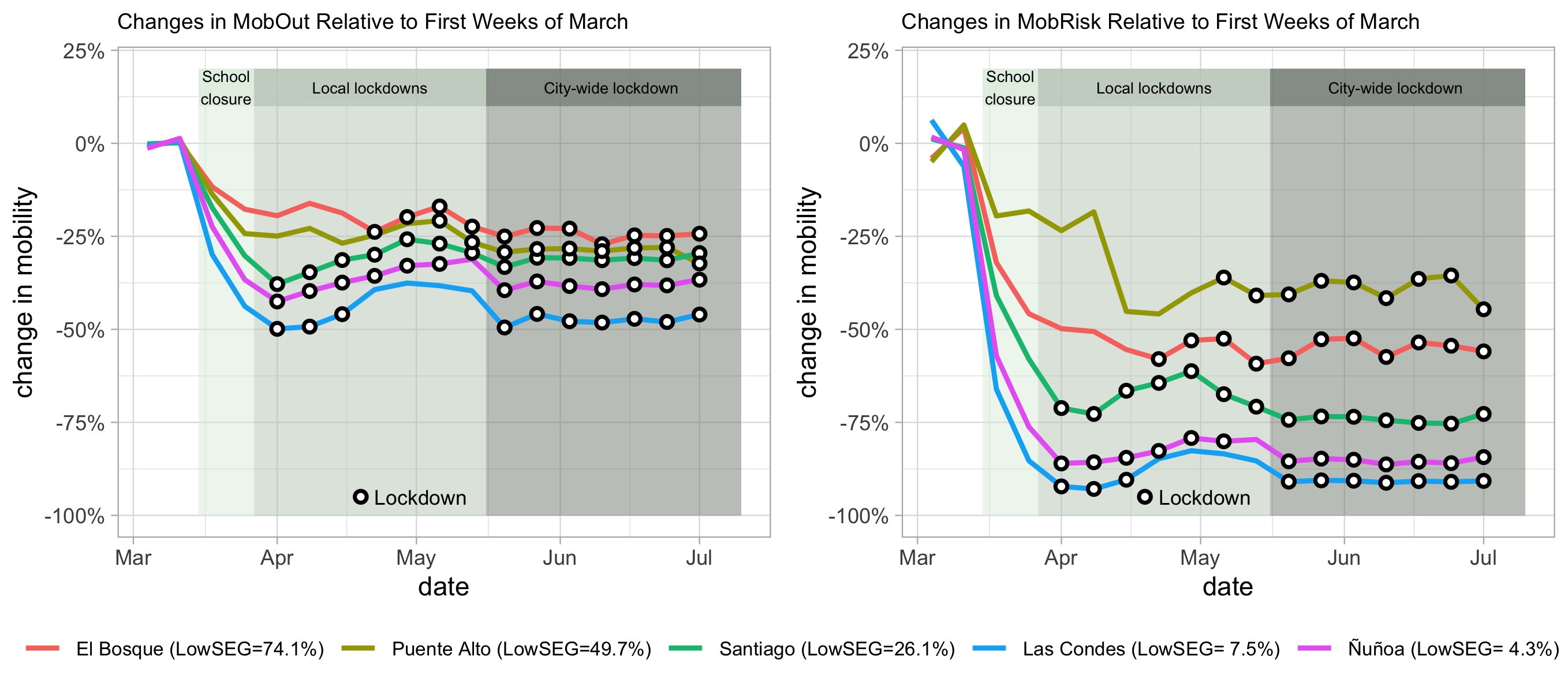}
    \caption{Temporal evolution of mobility in different municipalities of Santiago, measured as fraction of total trips leaving home zones relative to the baseline period (March 2--13). Legend shows in parentheses the fraction of population in the low socioeconomic segment (LowSEG) for each municipality.}
    \label{fig:Temporal_evolution_of_mobility}
\end{figure}}
Then, on March 27, mandatory lockdowns were initiated in the municipalities of Santiago, Ñuñoa, and Las Condes (as well as other municipalities; these correspond to group 1 in Fig. \ref{fig:cases_group}). Although this reduced mobility by an additional 5 to 8 percentage points in these municipalities, the decrease is relatively small compared to that recorded before the imposition of mandatory lockdowns.
In the last two weeks of April, the higher-income municipalities were released from mandatory lockdown, while  lockdowns were imposed on Puente Alto and El Bosque (as well as other municipalities; these correspond to group 2 in Fig. \ref{fig:cases_group}). In these latter cases, we do not observe a significant effect in the reduction of mobility associated with mandatory lockdowns. Indeed, Figure \ref{fig:Temporal_evolution_of_mobility} illustrates that in the last few weeks of April mobility in higher-income municipalities -- which were no longer under lockdown -- remained well below that of the others that were under mandatory lockdowns.\footnote{These patterns are consistent with results presented in other reports that used data from another mobile phone company and differently constructed mobility measures \citep{IndiceMovilidadUDD}.} 
The differences in mobility between municipalities of different socioeconomic strata remain even as the entire city went into mandatory lockdown from mid-May through June. The mobility measure $MobRisk$ presents a similar evolution and heterogeneity across municipalities of different socioeconomic levels to $MobOut$, as presented in the right panel of Figure~\ref{fig:Temporal_evolution_of_mobility}. Table \ref{tab:summ_stats} shows summary statistics of these two mobility measures at the municipality level. (See also Appendix C for additional summary statistics at municipality and census zone levels).


\begin{table}[!htbp] \centering 
\begin{tabular}{@{\extracolsep{5pt}}lccccccc} 
\\[-1.8ex]\hline 
\hline \\[-1.8ex]
Statistic & \multicolumn{1}{c}{N} & \multicolumn{1}{c}{Mean} & \multicolumn{1}{c}{St. Dev.} & \multicolumn{1}{c}{Min} & \multicolumn{1}{c}{Pctl(25)} & \multicolumn{1}{c}{Pctl(75)} & \multicolumn{1}{c}{Max} \\ 
\hline \\[-1.8ex] 
$MobOut$ & 411 & 0.529 & 0.064 & 0.326 & 0.508 & 0.572 & 0.721 \\ 
$MobRisk$ & 411 & 0.893 & 0.444 & 0.240 & 0.648 & 1.004 & 5.381 \\ 
\hline \\[-1.8ex] 
\end{tabular}
\caption{Summary statistics of municipality-level mobility measures.}
\label{tab:summ_stats}
\end{table} 

In the next sections we present an econometric analysis confirming the patterns observed in Figures \ref{fig:cases_group} and \ref{fig:Temporal_evolution_of_mobility}. 
First, in Section \ref{sec:mobility} we show that after controlling for other possible factors, the main factor that explains the differences in mobility patterns between the municipalities in Santiago is indeed socioeconomic status. Then, in Section \ref{sec:infections} we show that mobility is in fact an important explanatory variable of infections. 

\section{The impact of socioeconomic level on mobility} \label{sec:mobility}

 In this section we explore the heterogeneous impact of shelter-in-place behavior and mandatory lockdowns on mobility and how they are moderated by socioeconomic level. 



\subsection{Econometric model}

Our main model is a panel data linear regression model with cross-section $i$ defined by census zones analyzed in each week $t$, which fully exploits the cross-sectional variation at the most granular level. The model is estimated with the sample period from March 2 to July 3, 2020 (weeks 9 to 26 in the calendar year), in which the government implemented closing of schools, recommended sheltering-in-place, and adopted mandatory localized lockdowns and, later, a citywide lockdown in Santiago as explained above. To measure mobility we consider the two metrics we defined in Section \ref{sec:mob_meas}: \emph{MobOut}, which measures whether users leave their residences, and \emph{MobRisk}, which additionally measures whether users move to more risky areas. We use a log transformation that facilitates a relative interpretation of the coefficients and provides a better fit to the data. Thus, our main model to study mobility is defined as
\begin{align}
\log{(\text{Mobility}_{it})} &=\delta_{i}+\sum_{s\in\mathcal{S}} \tau^s_{t} \cdot D^s_{i} +  \sum_{s\in\mathcal{S}} \beta^s D^s_{i}\cdot LockLocal_{it} + u_{it} \label{eq:FracOut_wkhet} \,
\end{align}
where $\mathcal{S}=\{$HighSEG$, $MedSEG$, $LowSEG$\}$. The parameters $\delta_{i}$ are zone fixed effects and we omit the constant in the regression. The covariates $D_i^s$ are continuous variables specifying the fraction of the population in zone $i$ that belongs to each socioeconomic group (note that $\sum_{s\in\mathcal{S}} D_i^s=1$). Hence, the week-SEG coefficients $\tau_t^s$ (which are multiplied by the fraction of the population in that SEG) capture citywide interventions and other exogenous shocks on a given week that could potentially have {\em heterogeneous effects across SEGs}. For example, these week-SEG coefficients of the first weeks (weeks 11 and 12) capture voluntary shelter-in-place behavior at the beginning of the pandemic before the imposition of mandatory lockdowns.  Further, the coefficients of week 20 capture mobility reductions associated with the citywide lockdown, which started that week. We omit the week-SEG coefficients for the first week of March, so that all the effects are measured with respect to this ``pre-pandemic'' week. The coefficients of $LockLocal_{it}$ together with the week-SEG coefficients $\tau_t^s$, all of which are estimated separately for the different socioeconomic levels, are the main treatment effects to be estimated in this model. 
 
The treatment variable \emph{LockLocal} captures the effect of {\em localized lockdowns}. We construct this variable to represent the fraction of days in week $t$ that census zone $i$ is under a localized lockdown. The staggered implementation of lockdowns across zones with different demographic characteristics provides an exogenous variation that allow us to identify the effect of these localized lockdowns and their heterogeneous effect across zones. As discussed above, the week-SEG coefficients  capture the effect of the citywide lockdown, and so we set  the variables $LockLocal$ equal to zero during the city's full lockdown. 

We also run a parsimonious model that fixes the treatment effect and week coefficients  to be homogeneous across zones (presented in Appendix F along with other alternative specifications and robustness checks). The qualitative findings of these complementary analyses are similar to those derived from the main models.

\subsection{Estimation results}

The main estimation results correspond to the weekly-SEG coefficients  and the variation in mobility due to localized lockdowns. We start discussing the results for the week-SEG coefficients. To facilitate visualization, we present the estimates for each socioeconomic group in Figure \ref{fig:weekEstrato.modelFullInteraction} (the table with the parameter estimates are included in Appendix F). Here, the three lines represent the percentage variation in mobility for hypothetical zones with 100\% of their population in the corresponding group. 

\begin{figure}[hbt]
    \centering
    \includegraphics[width=.49\textwidth]{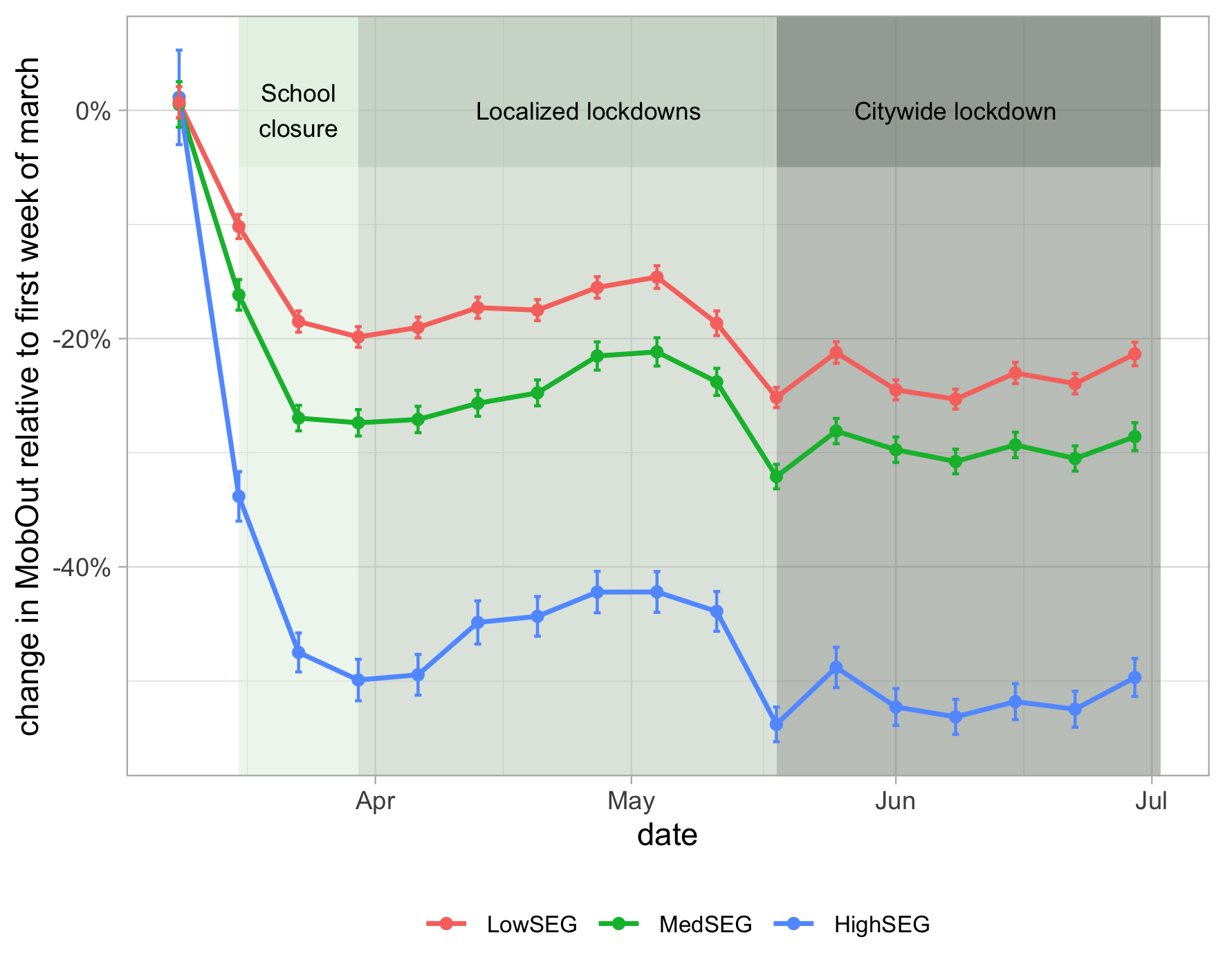}
    \includegraphics[width=.49\textwidth]{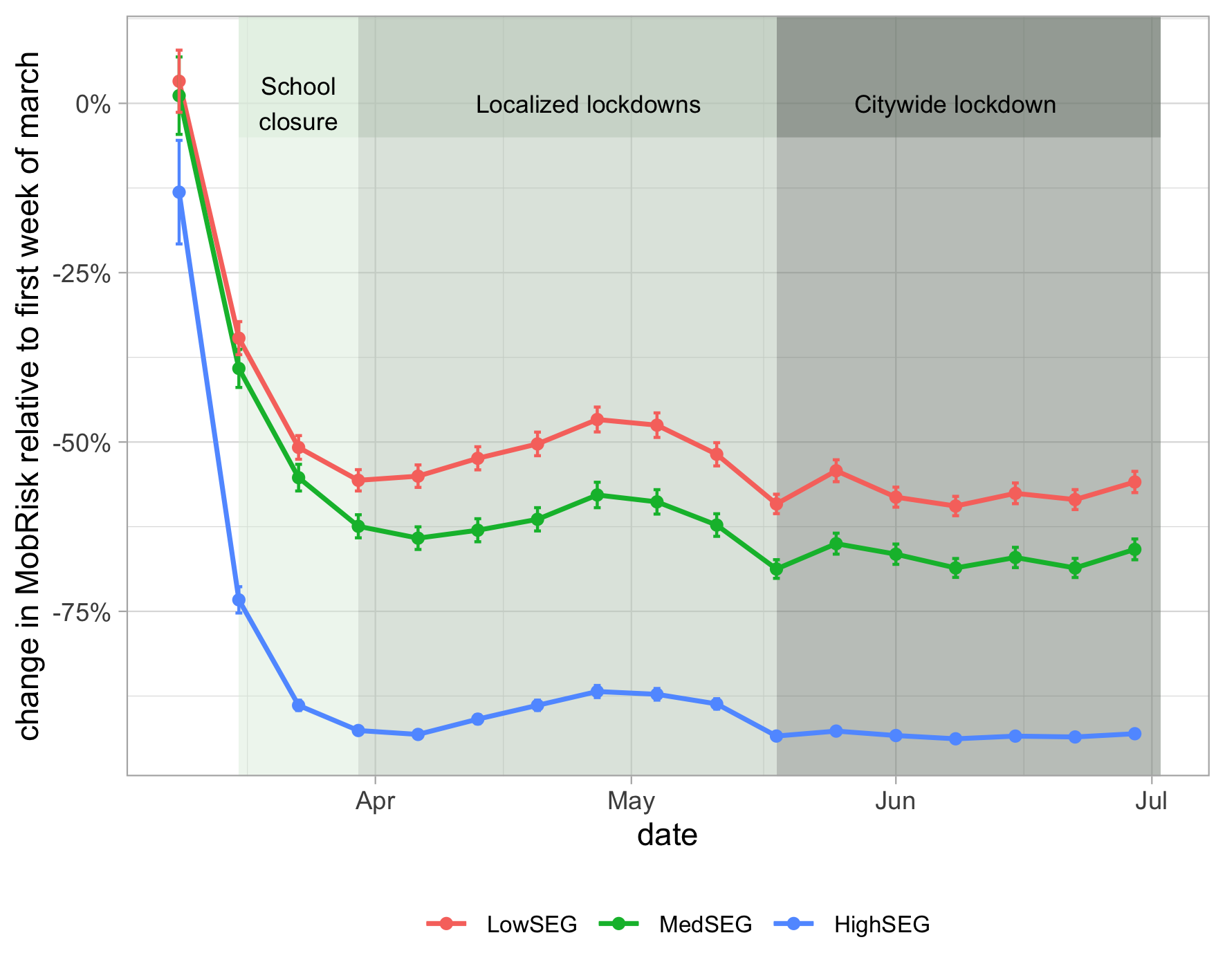}
    \caption{Week coefficients  interacted with socioeconomic group (week-SEG coefficients) for log(MobOut) and log(MobRisk) regressions. Error bars represent the 95\% confidence interval of the estimates (calculated using the Delta Method and robust standard errors).}
    \label{fig:weekEstrato.modelFullInteraction}
\end{figure}



Consider first the left panel of Figure 3 with the variation in \emph{MobOut}. The introduction of voluntary shelter-in-place directives and school closings caused a dramatic drop in mobility for the high socioeconomic group, on the order of 50\%. This is the largest effect we found across all the interventions analyzed in this study. The effect of voluntary restrictions is much smaller in the other groups: around 30\% for the medium and 20\% for the low socioeconomic group. These differences persisted during the period when localized lockdowns were enforced; note that the effect shown in the figure is after partialling out the effect of the localized lockdowns. The right panel displays week-SEG coefficients for \emph{MobRisk} and exhibits similar patterns, but the proportional change is larger in magnitude for this alternative mobility metric. The citywide lockdown was made effective on May 15, which is associated with an additional drop in both metrics of mobility from May 18 (week 20) onwards (the week-SEG coefficients  absorb the effect of the city-wide lockdown). 
Table \ref{tab:results_mob} displays parameter estimates for the coefficients associated with localized lockdowns. Recall that the dependent variable is in logarithm; hence the coefficients can be interpreted as the percentage reduction in mobility of these interventions.\footnote{The precise percentage effect can be calculated as  $\exp(\beta) - 1\%$, which in this case is similar to the coefficient estimates.}

\begin{table}[htbp] \centering
  \caption{Estimation results on the heterogeneous effect of localized lockdowns on mobility ($MobOut$ and $MobRisk$). All specifications include zone and week-SEG coefficients (not reported for brevity). Robust standard errors are reported in parentheses.}
  \label{tab:results_mob} 
  \footnotesize
\begin{tabular}{@{\extracolsep{5pt}}lcc} 
\\[-1.8ex]\hline 
\hline \\[-1.8ex] 
 & \multicolumn{2}{c}{\textit{Dependent variable:}} \\ 
\cline{2-3} 
\\[-1.8ex] & $\log(MobOut_{it})$ & $\log(MobRisk_{it})$ \\ 
\\[-1.8ex] & (1) & (2)\\ 
\hline \\[-1.8ex] 
 $LockLocal_{it}$:$HighSEG_i$ & $-$0.089$^{***}$ & $-$0.303$^{***}$ \\ 
  & (0.011) & (0.022) \\ 
  & & \\ 
 $LockLocal_{it}$:$MedSEG_i$ & $-$0.084$^{***}$ & $-$0.112$^{***}$ \\ 
  & (0.004) & (0.010) \\ 
  & & \\ 
 $LockLocal_{it}$:$LowSEG_i$ & $-$0.060$^{***}$ & $-$0.155$^{***}$ \\ 
  & (0.005) & (0.010) \\ 
  & & \\ 
\hline \\[-1.8ex] 
Observations & 29,214 & 29,214 \\ 
R$^{2}$ & 0.680 & 0.814 \\ 
Adjusted R$^{2}$ & 0.661 & 0.802 \\ 
\hline 
\hline \\[-1.8ex] 
  & \multicolumn{2}{r}{\textit{Note:} $^{*}$p$<$0.05; $^{**}$p$<$0.01; $^{***}$p$<$0.001} \\ 
\end{tabular} 
\end{table}

The point estimates suggest that localized lockdowns had a statistically significant effect in reducing mobility for all socioeconomic groups. However, in general, the reduction is larger in higher-income zones. For example, linear tests with F-stat reveal that the reduction in $MobOut$ in the HighSEG is larger than the corresponding reduction in the LowSEG (p-val<0.01). Similarly, the reduction in $MobRisk$ in the HighSEG is larger than the corresponding reduction in MedSEG and the LowSEG (p-val<0.01). The results show that the voluntary shelter-in-place directives, the localized lockdowns, and the citywide lockdown all generated higher reductions in mobility for the high socioeconomic group; the largest heterogeneity observed, however, was in the voluntary shelter-in-place directive. Overall, our results provide evidence that socioeconomic conditions affect the effectiveness of these interventions, even within the boundaries of a city.




\section{The impact of mobility on infections} \label{sec:infections}

The measures of mobility developed in this work are focused on estimating voluntary shelter-in-place behavior and lockdown compliance, which were the main non-pharmaceutical interventions used in Santiago to contain the outbreak. In this section, we show that these measures of mobility are indeed an important predictor of outbreaks, and therefore a relevant indicator for monitoring the spread of infections that can be useful for planning lockdowns and other interventions.

\subsection{Modeling infections}  

Epidemics are known to exhibit exponential growth and their transmissibility is monitored using different metrics. A commonly used measure to track the spread of infections is the effective reproduction number $R_e$, which represents the expected number of offsprings generated by an index case. Several methods have been developed to estimate $R_e$ using time-series data \citep{cori2013new,wallinga2004different}, some of which have been tailored specifically for Covid-19 \citep{bandt2020reproduction}. One limitation of these methods is that they require as an input the generation-time distribution (the time between the infection of the index case and its offsprings), which may vary depending on behavior, social distancing, and contact-tracing interventions \citep{gostic2020practical}). For our analysis, we use a simpler measure for the spread of the epidemic based on the growth rate of infections. More specifically, we utilize the ratio of concurrent infections with respect to infections in the previous period as a proxy for the reproduction rate of the virus. We use official data for infections, counted at the date of symptom onset, which for Santiago is reported weekly at the municipality level.  Let $y_{kt}$ be the number of infected individuals in municipality $k$ that began exhibiting symptoms in week $t$. Using this data, we define the growth rate measure at each municipality $k$ and week $t$ as $ R_{kt}=\frac{y_{kt}}{y_{k,t-1}}$. Note that this measure is only well defined for municipality $k$ in week $t$ in which $y_{k,t-1}>0$. Moreover, this measure may be imprecise in the initial weeks of the pandemic for municipalities that have a growing yet small number of cases. To deal with these issues, we consider only municipalities in weeks in which the corresponding cumulative number of cases exceeds a minimum threshold of 150 cases.\footnote{In the peak weeks, the weekly average number of new infections per municipality was around 1,000 cases.}

We use panel data linear regression models with cross-section $k$ defined by municipalities analyzed in each week $t$ between March 2 and July 3, 2020 (weeks 9 to 26) and include week dummies to capture citywide shocks in infections. Our first model measures the impact of mobility on infections through the regression
\begin{align} 
\log R_{kt} &= \delta_k + \tau_t +\theta_1\cdot \mbox{Mobility}_{k,t-1} + e_{kt} \ , \label{eq:inf_mob}
\end{align}
where $\delta_{k}$ is the municipality fixed effect,  $\tau_{t}$ are week dummy variables, and $e_{kt}$ is the error term. Mobility is measured using  \textit{MobOut} or \textit{MobRisk} in two alternative specifications; the main difference is that $MobRisk$ accounts for the spread of infections across municipalities to better account for infection risk, as described in Section \ref{sec:mob_meas}. The models use a one-week lag on the mobility variable, which is appropriate since the serial interval --the time between the symptom onset of the index case and its offsprings-- is on average 5--6 days for Covid-19 \citep{lauer2020incubation,rai2020estimates}. Note that because infections are measured at the municipality level, our mobility measures in this regression are also at this level of aggregation. The log transformation of the dependent variables is used to facilitate the interpretation of the mobility coefficients.\footnote{As with most studies analyzing Covid, the number of cases is usually under-reported because of asymptomatic cases. However, if the fraction of asymptomatic cases (from the total number of cases) is relatively constant, then the growth of the infections  is not subject to a major bias due to this under-reporting \citep{bandt2020reproduction}. There could also be under-reporting due to lack of testing capacity. However, our regression includes week fixed effects which absorbs weekly changes on testing capacity at the city level (which were quite large during the period under study). To further handle potential issues related to under-reporting due to insufficient testing capacity that could be different by municipalities, we used the method developed by \cite{russell2020reconstructing} using mortality data (see Appendix E). The results remain similar after this correction.}
 
The sample period comprises  the beginning of the outbreak (mid-March) until the beginning of July, which, as presented in Figure \ref{fig:cases_group}, includes a phase of voluntary shelter-in-place directives (until the end of March), localized lockdown (until mid-May), and then a citywide lockdown. The week fixed effects $\tau_t$ capture changes in behavior of the population, such as the use of hygiene measures, wearing masks, and avoiding crowded places. Our analysis in Section \ref{sec:mobility} shows an association between mandatory lockdowns and mobility, but it is plausible that lockdowns also lead to additional changes in behavior that help to reduce infections (other than forbearance from mobility). The effect of the citywide lockdown through these other behavioral factors is also captured through the week fixed effects. However, during the period of localized lockdowns, these behavioral factors become omitted variables that may lead to estimation bias, in the sense of overestimating the effect of forbearance of mobility on infections. To mitigate this problem, we specify an alternative model that includes localized lockdowns as a control variable:
\begin{align} 
\log{R_{kt}}&=\delta_{k}+\tau_{t}+\theta_1\cdot \mbox{Mobility}_{k,t-1} + \theta_2\cdot LockLocal_{k,t-1} + e_{kt}, \label{eq:mob_lock}
\end{align}
where $LockLocal_{kt}$ is the fraction of census zones in municipality $k$ that are under lockdown on every day of week $t$.\footnote{$LockLocal$ is similar to the localized lockdown variable used in regression \eqref{eq:FracOut_wkhet} but aggregated at the municipality level.}  In this model, the coefficient of \textit{LockLocal} should be interpreted as the effect of localized lockdowns on infections that is \textit{not} mediated by mobility and therefore attributed to other behavioral factors. It is still possible, however, that changes in mobility may also lead to other changes in behavior (not directly associated with lockdowns), and therefore, some of these behavioral adjustments will be absorbed by the Mobility measures.\footnote{\cite{jung2021} uses lockdowns as an instrumental variable for mobility. We think that lockdowns may not meet the exclusion restriction of an instrument if they  generate risk-awareness in the population, as described above.}


\subsection{Estimation results} \label{sec:inf_model_results}
    
The main estimation results are reported in Table \ref{tab:main_mobRt_results1} (week and municipality fixed effects are omitted in the table for brevity but included in the model). Columns (1) and (2) report the estimation results of model \eqref{eq:inf_mob} for the two measures of mobility: \textit{MobOut} and \textit{MobRisk}. Both measures are positive and statistically significant, suggesting a strong association between mobility and infections. Columns (3) and (4) correspond to the same model but restrict the sample to the period of citywide lockdown (mid-May to early July). Here, we observe that \textit{MobRisk} is positive and statistically significant, but \textit{MobOut} is no longer significant. That is, controlling for the citywide lockdown across all municipalities, our risk-adjusted mobility measure explains differences in infection rates across municipalities, whereas \textit{MobOut} -- a proxy for shelter-in-place compliance -- does not. As further validation, Columns (5) and (6) show the results of model \eqref{eq:mob_lock} during the entire sample period and after controlling for localized lockdowns (\textit{LockLocal}). Again, we see that the effect of \textit{MobRisk} is positive and statistically significant, but \textit{MobOut} is not significant. These results confirm the importance of considering the interaction between municipalities and regions \citep{holtz2020interdependence,akbarpour2020socioeconomic,birge2020controlling,fajgelbaum2020optimal,zubi2020} and suggest that, in order to capture the effects of mobility on infections, it is important to account for the patterns of population movements within the city and how these interact with local outbreaks. The results also highlight the importance of controlling for interventions that seek to induce social distancing in order to disentangle the effect of specific mobility measures on infection rates.

To evaluate the magnitude of the effect of \textit{MobRisk} on infections, we conduct a counterfactual analysis where the mobility of a low-income municipality is reduced to the mobility level of a high-income municipality. Using the data from Figure \ref{fig:Temporal_evolution_of_mobility}, panel (b), we evaluate the reduction in infections for the municipality with the highest \textit{MobRisk} (Puente Alto, which has a high fraction of low-income population) when reducing risk-adjusted mobility to match the level of the average between Las Condes and Ñuñoa, which have higher income and lower levels of \textit{MobRisk}.  The counterfactual is evaluated during the period when the citywide lockdown was active, so that all municipalities are subject to the same restrictions. This average reduction in mobility for the low-income municipality (Puente Alto) corresponds to 
67\%, and our analysis suggests that it 
 translates to an average  reduction in the infection rate of 36\% over the period considered (this is roughly half-point reduction in infection rate per every percentage point reduction in mobility).\footnote{
 The reduction in infections is calculated as the average of $\exp(\hat\theta_1\times \Delta MobRisk_t)-1$ over the time horizon, where $\Delta MobRisk_t$ is the difference of $MobRisk$ between the municipalities mentioned above in week $t$, and  $\hat\theta_1=0.252$ is the coefficient of $MobRisk$ from Table 3, column (6).}





Overall, these results provide robust evidence that reductions in mobility play a role in moderating the spread of the pandemic. Our analysis indicates that variations in mobility could potentially induce different infection rates even within a city, and therefore are a useful indicator for monitoring and controlling outbreaks. More specifically, our analysis shows that the stark differences in mobility observed across socioeconomic groups can explain the higher infection rates observed in lower-income urban areas.

\begin{table}[!htbp] 
    \centering
    \caption{Estimation results of the effect of mobility on infection rates ($\log(R_t)$) using an infection rate threshold of 150. All specifications include municipality and week fixed effects (not reported for brevity).}
\small 
\begin{tabular}{@{\extracolsep{5pt}}lcccccc} 
\\[-1.8ex]\hline 
\hline \\[-1.8ex] 
 & \multicolumn{6}{c}{\textit{Dependent variable: $log R_{kt}$}} \\ 
\cline{2-7} 
\\[-1.8ex] & (1) & (2) & (3) & (4) & (5) & (6)\\ 
\hline \\[-1.8ex] 
 $LockLocal_{k,t-1}$ &  &  &  &  & $-$0.209$^{***}$ & $-$0.153$^{***}$ \\ 
  &  &  &  &  & (0.058) & (0.048) \\ 
  & & & & & & \\ 
 $MobOut_{k,t-1}$ & 1.585$^{**}$ &  & $-$0.114 &  & $-$0.216 &  \\ 
  & (0.658) &  & (0.920) &  & (0.817) &  \\ 
  & & & & & & \\ 
 $MobRisk_{k,t-1}$ &  & 0.347$^{***}$ &  & 0.556$^{**}$ &  & 0.252$^{***}$ \\ 
  &  & (0.084) &  & (0.248) &  & (0.089) \\ 
  & & & & & & \\ 
\hline \\[-1.8ex] 
Observations & 411 & 411 & 204 & 204 & 411 & 411 \\ 
R$^{2}$ & 0.828 & 0.833 & 0.570 & 0.582 & 0.834 & 0.838 \\ 
Adjusted R$^{2}$ & 0.805 & 0.810 & 0.467 & 0.483 & 0.811 & 0.815 \\ 
\hline 
\hline \\[-1.8ex] 
  & \multicolumn{6}{r}{\textit{Note:} $^{*}$p$<$0.1; $^{**}$p$<$0.05; $^{***}$p$<$0.01} \\ 
\end{tabular} 
    \label{tab:main_mobRt_results1} 
\end{table}


\section{Discussion and policy implications}

We used detailed mobility data to understand the heterogeneous impact of lockdowns and voluntary shelter-in-place behavior in containing the Covid-19 pandemic within a large city of a developing country. Using granular data, we show that the effectiveness of interventions that seek to reduce mobility varies significantly depending on the socioeconomic level of the population.  More affluent zones exhibited large reductions in mobility due to voluntary shelter-in-place interventions in particular, but also due to mandatory lockdowns. By contrast, the compliance with these interventions was much lower in lower-income areas. We also show that reducing mobility has an impact on reducing the spread of the infections and is therefore an important indicator for monitoring outbreaks during a pandemic. Further, our detailed mobility data capturing origin-destination trips is useful for constructing risk-adjusted metrics that capture the agglomeration of infections in specific areas and the implied externalities that these have on infection transmission in the population.

There are a number of factors that might explain these different patterns of mobility across income groups. First, the high socioeconomic group presents a large share of professionals with the flexibility to work from home,  which can explain the large voluntary reductions in mobility for this group. Similarly, the effect of a lockdown might be larger in the high socioeconomic group because a larger fraction of their trips may be associated with non-working discretionary activities, whereas for other groups mobility is mostly driven by income needs. Further, our results show that even within a city, there are important dynamics at play to explain the effectiveness of policies for reducing mobility. 

Finally, our work highlights the challenges of reducing mobility in lower-income communities, where people generate their income from their daily work. To be effective, lockdowns and voluntary shelter-at-home directives have to be complemented by other measures that induce their inhabitants to increase compliance. Eventually, the Chilean government implemented two such measures: providing care packages of essential goods, as well as direct financial help to the most vulnerable communities. Preliminary anecdotal evidence suggests that these measures reduced mobility. Developing econometric analyses confirming this is an interesting topic for future research that could further illuminate how to effectively implement shelter-in-place policies in cities with socieconomic differences.

\subsection*{Acknowledgements}
We thank Ricardo Baeza-Yates, Eduardo Engel, Rafael Epstein, Diego Gil, Cristobal Huneeus, the journal editor, associate editor, and referees, and seminar participants from various conferences and institutions for their insightful comments. We thank Digital Entel Ocean for sharing valuable data. Financial support has partially come from ISCI (grant ANID PIA AFB180003). M. Goic also thanks the financial support of MIPP (IS130002). In addition, A. Carranza and G.Y. Weintraub thank the  Stanford RISE COVID-19 Crisis Response Faculty Seed Grant Program for helpful financial support.

\linespread{1.0}
\bibliographystyle{informs2014} 
\bibliography{main.bib}







\end{document}